\documentclass[aps,prd,twocolumn,showpacs,nofootinbib]{revtex4-2}

\usepackage{graphicx}
\usepackage{amsmath}
\usepackage{amssymb}
\usepackage{bm}
\usepackage{bbm}
\usepackage{color}
\usepackage{slashed}
\usepackage[colorlinks=true,linkcolor=blue,citecolor=blue, urlcolor=blue]{hyperref}

\begin{document}

\title{$Z$-boson decays into $S$-wave quarkonium plus a photon up to ${\cal O}(\alpha_{s} v^2)$ corrections}

\author{Guang-Yu Wang$^{a}$}
\email{wanggy@cqu.edu.cn}
\author{Xu-Chang Zheng$^{a}$}
\email{zhengxc@cqu.edu.cn}
\author{Xing-Gang Wu$^{a}$}
\email{wuxg@cqu.edu.cn}
\author{Guang-Zhi Xu$^{b}$}
\email{xuguangzhi@lnu.edu.cn}

\affiliation{$^a$ Department of Physics, Chongqing Key Laboratory for Strongly Coupled Physics, Chongqing University, Chongqing 401331, People's Republic of China \\
$^b$ Department of Physics, Liaoning University, Shenyang 110036, People's Republic of China}

\begin{abstract}

In this paper, we calculate the decay widths and branching fractions for the decays $Z \to H+ \gamma$ up to ${\cal O}(\alpha_{s} v^2)$ accuracy within the framework of nonrelativistic QCD, where $H$ stands for the $S$-wave quarkonium $\eta_c$, $J/\psi$, $\eta_b$ or $\Upsilon$, respectively. To compare with the leading-order terms, those corrections show good perturbative behavior as expected. It is found that contributions from the next-to-leading order QCD correction ${\cal O}(\alpha_{s}v^0)$, the relativistic correction ${\cal O}(\alpha^{0}_{s}v^2)$ and their joint correction ${\cal O}(\alpha_{s} v^2)$ are sizable and comparable to each other, especially for the charmonium case. Thus we need to take all of them into consideration for a sound estimation. For a high luminosity electron-positron collider running around the $Z$-pole, due to $Z$-boson resonance effect, sizable events could be produced from those rare decay channels.

\end{abstract}

\maketitle

\section{Introduction}
\label{secIntro}

Heavy quarkonium production involves both perturbative and nonperturbative aspects of QCD, it presents an ideal laboratory for studying the strong interaction. An effective theory, nonrelativistic QCD (NRQCD)~\cite{Bodwin:1994jh}, provides a rigorous way for separating the perturbative and nonperturbative effects in the heavy quarkonium production. In NRQCD, the quarkonium production rate is factorized as sums of the products of the short-distance coefficients (SDCs) and the long-distance matrix elements (LDMEs). SDCs are perturbatively calculable and can be expanded as a series over the strong coupling constant $\alpha_s$. LDMEs are nonperturbaive but universal, which have definite powers in $v_{Q}$ (the velocity of heavy quark $Q$ in the rest frame of heavy quarkonium). For simplicity, if not specially stated, we will always take $v\equiv v_{Q}$ throughout the paper. Then the NRQCD production rate can be represented as a double expansion over $\alpha_s$ and $v^2$. Till now, the NRQCD has achieved great successes in describing the heavy quarkonium production at different higher-energy colliders. However, there are still challenges to NRQCD, e.g., the polarization puzzle and LDME universality problem, cf. Refs.\cite{Brambilla:2010cs, Andronic:2015wma, Lansberg:2019adr, Chapon:2020heu, Chen:2021tmf}. It is important to study more production channels so as to understand the heavy quarkonium production mechanism.

Among the heavy quarkonium production channels, the exclusive decays of the $Z$-boson to a quarkonium and a photon are of special interest. These exclusive decay processes have clean experimental signatures, which can be easily distinguished even in the challenging environment of a hadron collider. They provide good platforms for studying the properties of the heavy quarkonium. Since the charge parity is conserved in the electromagnetic and strong interactions, the production of a heavy quakonium plus a photon is through either the vector component or the axial-vector component of the $Z$ boson. Therefore, these decay processes can also be used to study the $Z$ boson couplings to the heavy quarks \cite{Dong:2022ayy}. Furthermore, the experimental study of these decay processes can refine the experimental techniques on measuring the decay rates of the Higgs-boson decays into a heavy quarkonium and a photon \cite{Huang:2014cxa, Grossman:2015cak}, which can be inversely adopted to determine the Higgs-boson couplings to the charm and bottom quarks accordingly~\cite{Bodwin:2013gca}.

The experimental searches for the $Z$-boson decays into a vector heavy quarkonium and a photon have been carried out by the ATLAS and CMS collaborations at the LHC \cite{ATLAS:2015vss, ATLAS:2018xfc, CMS:2018gcm, ATLAS:2022rej}, but no significant signals have been observed. In recent years, several high-luminosity $e^+e^-$ colliders have been proposed, such as ILC \cite{ILC:2013jhg}, FCC-ee \cite{FCC:2018evy}, CEPC \cite{CEPCStudyGroup:2018ghi}, and Super $Z$ factory \cite{zfactory}. They are planed to run at the $Z$ pole for a period of time, thus a large number of $Z$-boson events are expected to be generated. These proposed $e^+e^-$ colliders provide new opportunities for studying the rare decays of the $Z$ boson.

The first theoretical study for $Z$-boson decays into a photon and a quarkonium was carried out by the authors of Ref.\cite{Guberina:1980dc} in 1980. The authors there calculated the decay rates at leading order (LO) in $\alpha_s$ and $v^2$. In Ref.\cite{Luchinsky:2017jab}, the $Z$-boson decays into a photon plus an $S$-wave or $P$-wave charmonium were studied at LO in $\alpha_s$ and $v^2$ using both the NRQCD and light-cone distribution amplitude (LCDA) approaches. In Ref.\cite{Wang:2013ywc}, the analytic expressions for the amplitudes of these decays up to next-to-leading order (NLO) in $\alpha_s$ in the LCDA approach were presented. In Ref.\cite{Huang:2014cxa}, the decays $Z\to V+\gamma$ where $V$ denotes a vector heavy quarkonium $J/\psi$ or $\Upsilon$ were calculated up to NLO in $\alpha_s$ and $v^2$ in both the NRQCD and LCDA approaches. In addition to the direct contribution, the authors of Ref.\cite{Huang:2014cxa} also calculated the indirect contribution where the $Z$-boson decays to a real photon and a virtual photon through a heavy-quark or $W$-boson loop with the virtual photon fragmenting into a heavy quarkonium. In Ref.\cite{Grossman:2015cak}, the decays $Z \to V+\gamma$ were studied in the LCDA approach, where the leading logarithms (LL) of $m_{_Z}^2/m_Q^2$ appearing in the perturbative series were resummed. In Ref.\cite{Bodwin:2017pzj}, the large logarithms of $m_{_Z}^2/m_Q^2$ were resummed to next-to-leading-logarithmic (NLL) accuracy. In Ref.\cite{Dong:2022ayy}, the decays $Z\to \Upsilon(nS)+\gamma$ were proposed to be used to probe the $Z b\bar{b}$ anomalous couplings, and the decay rates were calculated up to NLO in $\alpha_s$ in the NRQCD approach. Recently, the decay rates for the $Z$-boson decays into a photon plus an $S$-wave or $P$-wave quarkonium were studied up to next-to-next-to-leading order (NNLO) in $\alpha_s$ in the NRQCD approach \cite{Sang:2022erv, Sang:2023hjl}. Besides, the exclusive production processes of a photon plus a heavy quarkonium from $e^+e^-$ collisions \cite{Chung:2008km, Li:2009ki, Sang:2009jc, Braguta:2010mf, Fan:2012dy, Gao:2013jrw, Chen:2013itc, Li:2013nna, Sun:2014kva, Xu:2014zra, Brambilla:2017kgw, Chung:2019ota, Sang:2020fql,
Liao:2021ifc, Bhatnagar:2023mvj} and Higgs-boson decays \cite{Bodwin:2013gca, Bodwin:2014bpa, Bodwin:2016edd, Zhou:2016sot, Modak:2016cdm, Bodwin:2017wdu, Mao:2019hgg, Brambilla:2019fmu}, which have the same final states as the decays $Z \to H + \gamma$, have also been studied extensively in references.

In the present paper, we devote ourselves to calculating the decay rates of the $Z$-boson decays into an $S$-wave heavy quarkonium plus a photon up to ${\cal O}(\alpha_s v^2)$ corrections within the framework of NRQCD. The contributions from the relativistic ${\cal O}(\alpha_s^0 v^2)$ correction and the joint relativistic and radiative ${\cal O}(\alpha_s v^2)$ correction are expected to be comparable to QCD ${\cal O}(\alpha_s v^0)$ correction. Thus, they are important contents for the accurate prediction of the decay rates of these $Z$ decay processes.

The rest of the paper is organized as follows. In Sec.\ref{sec2}, we give the factorization formulism for the $Z$-boson decays to a photon plus an $S$-wave heavy quarkonium under the NRQCD framework. In Sec.\ref{sec3}, we sketch the method used in the perturbative QCD calculation. In Sec.\ref{sec4}, we present the numerical results and discussions. Section \ref{sec4} is reserved as a summary.

\section{NRQCD factorization formalism for the decay widths}
\label{sec2}

According to the NRQCD factorization, the amplitude up to order $v^2$ for the $Z$-boson decay to an $S$-wave heavy quarkonium and a photon can be factorized as
\begin{equation}
{\cal M}_{Z \to H+\gamma}=\left(c_0+c_2 \langle v^2 \rangle_H \right)\sqrt{2m_{_H}}\langle H \vert \psi^\dagger {\cal K} \chi \vert 0 \rangle,
\label{eq.nrqcdfact}
\end{equation}
where $c_0$ and $c_2$ are SDCs that can be expanded perturbatively in $\alpha_s$ as $c_i=c_i^{(0)}+c_i^{(1)}\frac{\alpha_s}{\pi}+\cdots$. $\langle H \vert \psi^\dagger {\cal K} \chi \vert 0 \rangle$ is the LDME, where ${\cal K}=1$ for the pseudo-scalar states $\eta_c$ ($\eta_b$) and ${\cal K}={\bm \sigma}\cdot {\bm \epsilon}$ for the vector states $J/\psi$ ($\Upsilon$), respectively, $\psi$ and $\chi$ are Pauli spinor fields that describe heavy quark annihilation and heavy antiquark creation, respectively. The factor $\sqrt{2m_{_H}}$ originates from that we adopt relativistic normalization for the quakonium state in ${\cal M}_{Z \to H+\gamma}$, but we adopt conventional nonrelativistic normalization for the NRQCD matrix element on the right-hand side of Eq.(\ref{eq.nrqcdfact}). The quantity $\langle v^2 \rangle_H$ is defined as
\begin{equation}
\langle v^2 \rangle_H=\frac{\langle H \vert \psi^\dagger \left(-\frac{i}{2}\tensor{\bf D}\right)^2{\cal K} \chi \vert 0 \rangle}{m_Q^2 \langle H \vert \psi^\dagger {\cal K} \chi \vert 0 \rangle},
\end{equation}
where the covariant derivative operator is defined as $\psi^\dagger \tensor{\bf D}\chi\equiv \psi^\dagger {\bf D}\chi-({\bf D} \psi)^\dagger \chi$.

Having the amplitude, the decay widths can be calculated through
\begin{equation}
\Gamma_{Z\to H+\gamma}=\frac{1}{3} \frac{1}{2 m_{_Z}}\frac{m_{_Z}^2-m_{_H}^2}{8\pi\, m_{_Z}^2} \sum \vert {\cal M}_{Z \to H+\gamma} \vert^2,
\label{eq.gamma}
\end{equation}
where $1/3$ comes from the average over the $Z$-boson polarizations, and $\sum$ means that all the polarizations of the initial and final state bosons have been summed up.

\section{Details for the perturbative QCD calculation}
\label{sec3}

In order to derive the first two SDCs $c_0$ and $c_2$, we first apply the NRQCD factorization to an on-shell $(Q\bar{Q})$ pair with proper quantum numbers, i.e.,
\begin{equation}
{\cal M}_{Z \to (Q\bar{Q})[n]+\gamma}=\left(c_0+c_2 \, v^2  \right)\langle (Q\bar{Q})[n] \vert \psi^\dagger {\cal K} \chi \vert 0 \rangle,
\label{eq.nrqcdfact-QQbar}
\end{equation}
where $n$ denotes the quantum numbers of $(Q\bar{Q})$ pair, i.e., $n=\,^1S_0^{[1]}$ for $\eta_c$ and $\eta_b$; $n=\,^3S_1^{[1]}$ for $J/\psi$ and $\Upsilon$. Here, the relativistic normalization is adopted for the NRQCD matrix element of the $(Q\bar{Q})$ pair on the right-hand side of Eq.(\ref{eq.nrqcdfact-QQbar}). The amplitude ${\cal M}_{Z \to (Q\bar{Q})[n]+\gamma}$ can be calculated perturbatively in QCD, and the NRQCD matrix element $\langle (Q\bar{Q})[n]\vert \psi^\dagger {\cal K} \chi \vert 0 \rangle$ can be calculated perturbatively in NRQCD. Then the SDCs can be extracted through comparing both sides of Eq.(\ref{eq.nrqcdfact-QQbar}). Since the SDCs are insensitive to the long-distance dynamics, the SDCs determined from the on-shell $(Q\bar{Q})$ pair production can be applied to the quarkonium production. In this work, the SDCs $c_0$ and $c_2$ are calculated up to NLO in $\alpha_s$.

\begin{figure}[htbp]
\includegraphics[width=0.45\textwidth]{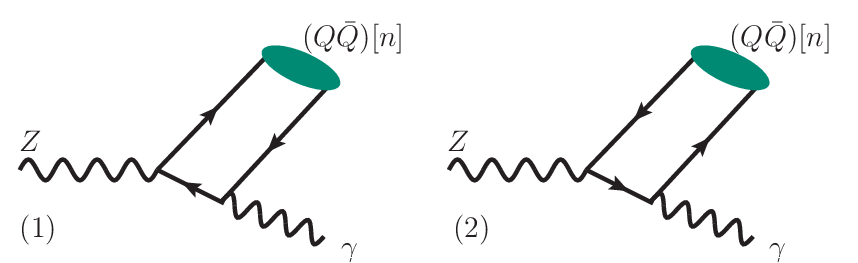}
\caption{Tree-level Feynman diagrams for the decay $Z\to (Q \bar{Q})[n] + \gamma$.
 } \label{feylo}
\end{figure}

\begin{figure}[htbp]
\includegraphics[width=0.45\textwidth]{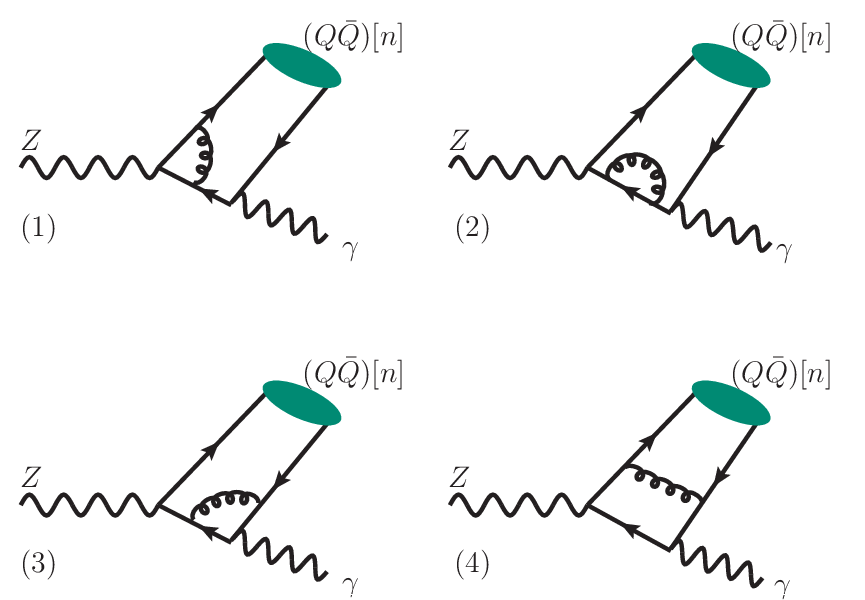}
\caption{Half of the one-loop Feynman diagrams for the decay $Z\to (Q \bar{Q})[n] + \gamma$.
 } \label{feynlo}
\end{figure}

There are two tree-level Feynman diagrams and eight one-loop Feynman diagrams for the decay $Z\to (Q \bar{Q})[n] + \gamma$. Two tree-level Feynman diagrams are shown in Fig.\ref{feylo}, and half of the eight one-loop Feynman diagrams are shown in Fig.\ref{feynlo}. The other four one-loop Feynman diagrams can be obtained from the four diagrams shown in Fig.\ref{feynlo} through reversing the fermion lines.

We assign the momenta of quark and antiquark in the $(Q\bar{Q})$ pair as $p_1=p+q$ and $p_2=p-q$, respectively. Then we have the following relations between the momenta:
\begin{eqnarray}
&& p_1^2=p_2^2=m_Q^2, ~p^2=E^2, ~p\cdot q=0,\nonumber \\
&& q^2=m_Q^2-E^2=-m_Q^2 v^2.
\end{eqnarray}
In the rest frame of $(Q\bar{Q})$ pair, we have $p=(E,0)$ and $q=(0,{\bm q})$, where $E=\sqrt{{\bm q}^2+m_Q^2}$.

It is convenient to use the covariant projection method to carry out the contributions of the different spin and color states. The spin projectors that are valid to all orders in $v$ have been derived in Refs.\cite{Bodwin:2002cfe,Bodwin:2010fi}. The projector for the spin-singlet state is
\begin{eqnarray}
\Pi_1=\frac{1}{2\sqrt{2}E(E+m_Q)}(\slashed{p}_2-m_Q)\gamma_5(\slashed{p}+E)(\slashed{p}_1+m_Q),
\end{eqnarray}
and the projector for the spin-triplet state is
\begin{eqnarray}
\Pi_3=\frac{1}{2\sqrt{2}E(E+m_Q)}(\slashed{p}_2-m_Q)\slashed{\epsilon}^*(\slashed{p}+E)(\slashed{p}_1+m_Q),
\end{eqnarray}
where $\epsilon^*$ denotes the polarization vector of the spin-triplet state. The projector for color-singlet state is
\begin{eqnarray}
\Lambda_1=\frac{\bf 1}{\sqrt{3}},
\end{eqnarray}
where ${\bf 1}$ denotes the unit color matrix. Then the amplitudes for the $(Q\bar{Q})$ pair production can be calculated through
\begin{eqnarray}
{\cal M}_{Z \to (Q\bar{Q})[n]+\gamma}={\rm Tr}(\widetilde{{\cal M}}\, \Pi_{1(3)} \Lambda_1),
\end{eqnarray}
where $\widetilde{{\cal M}}$ denotes the amplitudes for $Z \to (Q\bar{Q})[n]+\gamma$ with the external quark and antiquark spinors amputated. The trace acts on both the Dirac and color matrices.

To expand the amplitudes in $v$, we first expand the amplitudes as
\begin{eqnarray}
{\cal M}_{Z \to (Q\bar{Q})[n]+\gamma}=&& {\cal M}\Big{|}_{q=0}+q^{\alpha} \frac{\partial {\cal M}}{\partial q^{\alpha}}\Big{|}_{q=0}\nonumber \\
 && +\frac{1}{2!}q^{\alpha}q^{\beta}\frac{\partial^{2}{\cal M}}{\partial q^{\alpha} \partial q^{\beta}}\Big{|}_{q=0} +\cdots,
\label{eq.expand1}
\end{eqnarray}
where ${\cal M}_{Z \to (Q\bar{Q})[n]+\gamma}$ has been abbreviated as ${\cal M}$ on the right hand side of Eq.(\ref{eq.expand1}) for simplicity. Then we average over the spatial direction of the momentum $q$ in the rest frame of $(Q\bar{Q})$ pair using the relations
\begin{eqnarray}
\int \dfrac{d\Omega_{\hat{\bm q}}}{4\pi}q^{\alpha} =0, ~ ~ ~
\int\!\dfrac{d\Omega_{\hat{\bm q}}}{4\pi}q^{\alpha} q^{\beta} =\frac{{\bm q}^2}{D-1}\Pi^{\alpha\beta},
\end{eqnarray}
where $D$ denotes the number of the space-time dimensions, ${\bm q}^2=m_Q^2 v^2$, and
\begin{eqnarray}\label{eq_piexd}
\Pi_{\alpha\beta}\equiv-g_{\alpha\beta}+\frac{p_{\alpha}p_{\beta}}{p^2}.
\end{eqnarray}
Therefore, the expansion of the amplitudes up to relative order $v^2$ after averaging over the direction of ${\bm q}$ can be expressed as
\begin{eqnarray}
{\cal M}_{Z \to (Q\bar{Q})[n]+\gamma}=&& {\cal M}\Big{|}_{q=0}+\frac{1}{2!}\frac{{\bm q}^2\Pi_{\alpha\beta}}{(D-1)}\frac{\partial^{2}{\cal M}}{\partial q^{\alpha} \partial q^{\beta}}\Big{|}_{q=0}.\nonumber \\
\label{eq.expand2}
\end{eqnarray}

In the calculation, we expand the amplitudes in $v$ before performing the loop integration, this means that we calculate the contributions of the hard region in the language of the region method \cite{Beneke:1997zp}. Using this strategy, the Coulomb divergences which come from the potential region do not appear in our calculation.

There are ultraviolet (UV) and infrared (IR) divergences in the one-loop calculation. We adopt the dimensional regularization with $D=4-2\epsilon$ to regularize these divergences. Under the dimensional regularization, the UV and IR divergences appear as pole terms in $\epsilon$. The UV divergences need to be removed through renormalization. In the calculation, the renormalization for the heavy quark field and the heavy quark mass are carried out in the on-mass-shell (OS) scheme, and the renormalization constants are
\begin{eqnarray}
&&\delta Z_2^{\rm OS}=-\dfrac{C_F \alpha_s}{4\pi}\left[\dfrac{1}{\epsilon_{\rm UV}} +\dfrac{2}{\epsilon_{\rm IR}} -3\gamma_E +3\ln\dfrac{4\pi \mu_R^2}{m_Q^2} +4 \right], \nonumber \\
&&\delta Z_m^{\rm OS}=-\dfrac{3C_F \alpha_s}{4\pi}\left[\dfrac{1}{\epsilon_{\rm UV}} -\gamma_E +\ln\dfrac{4\pi \mu_R^2}{m_Q^2} +\frac{4}{3}\right],
\end{eqnarray}
where $C_F=4/3$, $\gamma_E=0.557$ is the Euler constant and $\mu_R$ is the renormalization scale. The IR divergences canceled in ${\cal M}_{Z \to (Q\bar{Q})[n]+\gamma}$ at order ${\cal O}(\alpha_s v^0)$. However, there are IR divergences remaining in ${\cal M}_{Z \to (Q\bar{Q})[n]+\gamma}$ at order ${\cal O}(\alpha_s v^2)$. As shown in the following, these IR divergences also appear in the renormalization of the LCDEs $\langle (Q\bar{Q})[n] \vert \psi^\dagger {\cal K} \chi \vert 0 \rangle$. Thus, the net contribution of the matrix element do not contain IR divergences.

The LDME $\langle (Q\bar{Q})[n] \vert \psi^\dagger {\cal K} \chi \vert 0 \rangle$ for an on-shell quark pair can be calculated perturbatively through its operator definition. Up to order ${\cal O}(\alpha_{s} v^2)$, we have \cite{Li:2013otv}
\begin{eqnarray}
&&\langle (Q\bar{Q})[n] \vert \psi^\dagger {\cal K} \chi \vert 0 \rangle_{\overline{\rm MS}}\nonumber \\
&&= \Bigg\{1+\frac{2\alpha_s C_F}{3 \pi}\left(\frac{\mu_R^2}{\mu_{\Lambda}^2}\right)^{\epsilon}\bigg[\frac{1}{\epsilon_{\rm IR}}+{\rm ln}(4\pi)-\gamma_E  \bigg] \frac{{\bm q}^2}{m_Q^2}\Bigg\}\nonumber \\
&&~~~   \times \langle (Q\bar{Q})[n] \vert \psi^\dagger {\cal K} \chi \vert 0 \rangle^{(0)},
\end{eqnarray}
where the subscript ``$\overline{\rm MS}$" indicates that the UV divergences in the LDME are removed through the renormalization under the $\overline{\rm MS}$ scheme, $\mu_{\Lambda}$ is the factorization scale for the LDME. $\langle (Q\bar{Q})[n] \vert \psi^\dagger {\cal K} \chi \vert 0 \rangle^{(0)}$ denotes the LO LDME, and we have
\begin{eqnarray}
&& \vert \langle (Q\bar{Q})[^1S_0] \vert \psi^\dagger {\cal K} \chi \vert 0 \rangle^{(0)}\vert^2=2N_c(2E)^2, \nonumber \\
&& \vert \langle (Q\bar{Q})[^3S_1] \vert \psi^\dagger {\cal K} \chi \vert 0 \rangle^{(0)}\vert^2=2N_c(D-1)(2E)^2,
\end{eqnarray}

The $\gamma_5$ matrix should be noted in dimensional regularization. In this work, we adopt the Larin scheme~\cite{Larin:1993tq}, in which the product of a $\gamma_{\mu=(1,\cdots,4)}$ matrix and a $\gamma_5$ matrix is expressed as
\begin{eqnarray}
\gamma_\mu\gamma_5=i\frac{1}{6}\epsilon_{\mu\rho\sigma\tau}\gamma^\rho\gamma^\sigma\gamma^\tau.
\end{eqnarray}
In this $\gamma_5$ scheme, finite renormalization should be introduced to the axial vector vertex to restore the axial current Ward identity. The corresponding renormalization constant $Z_5$ for the axial vector vertex is
\begin{eqnarray}
Z_5=1-\frac{\alpha_s}{\pi}C_F.
\end{eqnarray}

In the calculation, the FeynArts package \cite{feynarts} is used to generate Feynman diagrams and amplitudes, the Feyncalc package \cite{feyncalc1,feyncalc2} is used to carry out the trace over the color and Dirac matrices, and the \$Apart package \cite{apart} and the FIRE package \cite{fire} are used to reduce the one-loop integrals to the master integrals, and the master integrals are numerically calculated by using the package LoopTools \cite{looptools}.

\section{Numerical results and discussions}
\label{sec4}

In doing the numerical calculation, we adopt the following values for the input parameters:
\begin{eqnarray}
&& m_c=1.4\pm 0.1 \,{\rm GeV},\; m_b=4.6 \pm 0.1\,{\rm GeV},\; \nonumber \\
&& m_{_Z}=91.1876\,{\rm GeV}, \;\, {\rm sin}^2\theta_W=0.231, \nonumber \\
&& \alpha=1/128,
\label{eq.input}
\end{eqnarray}
where $m_c$ and $m_b$ are pole masses, $\alpha$ is the electromagnetic coupling constant at $m_{_Z}$. For the strong coupling constant, we adopt the two-loop formula
\begin{eqnarray}
\alpha_s(\mu_R)=\frac{4\pi}{\beta_0 L}\left( 1-\frac{\beta_1 {\rm ln} L}{\beta_0^2 L} \right),
\end{eqnarray}
where $L={\rm ln}(\mu_R^2/\Lambda_{\rm QCD}^2)$, $\beta_0=11-2n_f/3$, $\beta_1=102-38 n_f/3$, and $n_f$ denotes the number of active quark flavors. According to $\alpha_s(m_{_Z})=0.118$ \cite{ParticleDataGroup:2022pth}, we obtain $\Lambda^{n_f=5}_{\rm QCD}= 0.226\, {\rm GeV}$ and $\Lambda^{n_f=4}_{\rm QCD}= 0.327\, {\rm GeV}$.

We take the two color-singlet LDMEs $\langle \mathcal{O}_{1}\rangle_{\eta_c}=\vert \langle \eta_c \vert \psi^{\dagger}\chi \vert 0 \rangle \vert^2 $ and $\langle \mathcal{O}_{1}\rangle_{J/\psi}=\frac{1}{3}\sum_{\lambda}\vert \langle J/\psi((\lambda)\vert \psi^\dagger {\bm \sigma}\cdot {\bm \epsilon} (\lambda)\chi \vert 0 \rangle \vert^2$ for $\eta_c$ and $J/\psi$, and the quantities $\langle v^2\rangle_{\eta_c}$ and $\langle v^2\rangle_{J/\psi}$ from Ref.\cite{Bodwin:2007fz}, i.e.,
\begin{eqnarray}
&&\langle \mathcal{O}_{1}\rangle_{\eta_c}=0.434^{+0.169} _{-0.158}\,{\rm GeV}^{3}, \\
&&\langle \mathcal{O}_{1}\rangle_{J/\psi}=0.440^{+0.067} _{-0.055} \,{\rm GeV}^{3},
\label{eq.ldmes1}
\end{eqnarray}
and
\begin{eqnarray}
&&\langle v^2\rangle_{\eta_c}=0.226^{+0.123} _{-0.098}, \\
&&\langle v^2\rangle_{J/\psi}=0.225^{+0.106} _{-0.088}.
\label{eq.ldmes2}
\end{eqnarray}
We take the LDME $\langle \mathcal{O}_{1}\rangle_{\Upsilon }$ and the quantity $\langle v^2\rangle_{\Upsilon}$ for $\Upsilon$ from Ref.\cite{Chung:2010vz}, i.e.,
\begin{eqnarray}
&&\langle \mathcal{O}_{1}\rangle_{\Upsilon }=3.069^{+0.207} _{-0.190}\,{\rm GeV}^{3}, \\
&&\langle v^2\rangle_{\Upsilon}=-0.009^{+0.003} _{-0.003}.
\label{eq.ldmes3}
\end{eqnarray}
As for the $\eta_b$ case, we adopt the approximation relations $\langle \mathcal{O}_{1}\rangle_{\eta_b} \approx \langle \mathcal{O}_{1}\rangle_{\Upsilon }$ and $\langle v^2\rangle_{\eta_b} \approx \langle v^2\rangle_{\Upsilon}$. Those values indicate that the spin splitting effects are small for heavy quarkonia, especially for the case of bottomium.

\subsection{Basic results}

In this subsection, we present numerical results for the decay widths of $Z \to H+ \gamma$ up to ${\cal O}(\alpha_{s} v^2)$, where $H =\eta_{c}$, $J/\psi$, $\eta_{b}$ or $\Upsilon$, respectively. To have a glance on the size of the contributions from different terms, we first present the decay widths when the quark masses and the LDMEs are taken as their central values. An analysis of the uncertainties will be presented next subsection.

\begin{table}[htb]
\begin{tabular}{c c c c c c}
\hline\hline
$\alpha_s(\mu_R)$  &  ${\rm LO}$  & ${\cal O}(\alpha_s)$ & ${\cal O}(v^2)$ & ${\cal O}(\alpha_{s} v^2)$ & {\rm Total} \\
\hline
$\alpha_s(2 m_c) =0.263$             &  $40.04$    &  $8.31   $  & $-7.55$ & $-7.50$   &$ 33.31  $ \\
$\alpha_s({m_{_Z}}/{2}) =0.132$  &  $40.04 $   &  $4.17   $  & $-7.55$  & $-3.77$ & $ 32.90  $\\
$\alpha_s(m_{_Z}) =0.118$             &  $40.04 $   &  $3.73   $  & $-7.55$  & $-3.37$ & $ 32.86  $\\
\hline\hline
\end{tabular}
\caption{Total and separate decay widths (in unit: eV) of $Z \to \eta_c + \gamma $ up to order ${\cal O}(\alpha_{s} v^2)$.}
\label{tb.etac}
\end{table}

\begin{table}[htb]
\begin{tabular}{c c c c c c}
\hline\hline
$\alpha_s(\mu_R)$  &  ${\rm LO}$  & ${\cal O}(\alpha_s)$ & ${\cal O}(v^2)$ & ${\cal O}(\alpha_{s} v^2)$ & {\rm Total} \\
\hline
$\alpha_s(2 m_c) =0.263$     & $ 275.58 $   &  $ 15.00 $  & $ -10.26 $  & $ -23.58 $ & $ 256.75  $ \\
$\alpha_s({m_{z}}/{2}) =0.132$  & $275.58  $   &  $7.53   $   & $-10.26  $  & $ -11.83 $ & $ 261.02   $\\
$\alpha_s(m_{z}) =0.118$     & $ 275.58 $   &  $ 6.73  $  & $ -10.26 $  & $ -10.58 $ & $ 261.48   $\\
\hline\hline
\end{tabular}
\caption{Total and separate decay widths (in unit: eV) of $Z \to J/\psi + \gamma $ up to order ${\cal O}(\alpha_{s} v^2)$.}
\label{tb.Jpsi}
\end{table}

\begin{table}[htb]
\begin{tabular}{c c c c c c}
\hline\hline
$\alpha_s(\mu_R)$  &  ${\rm LO}$  & ${\cal O}(\alpha_s)$ & ${\cal O}(v^2)$ & ${\cal O}(\alpha_{s} v^2)$ & {\rm Total} \\
\hline
$\alpha_s(2 m_b) =0.178$              &  $ 69.32 $ &  $-0.61  $  & $0.53$  & $0.21 $ & $ 69.45$ \\
$\alpha_s({m_{z}}/{2}) =0.132$   &  $ 69.32 $ &  $-0.46  $  & $0.53$  & $0.16 $ & $ 69.55 $\\
$\alpha_s(m_{z}) =0.118$              &  $ 69.32 $ &  $-0.41  $  & $0.53$  & $0.14 $ & $ 69.58 $\\
\hline\hline
\end{tabular}
\caption{Total and separate decay widths (in unit: eV) of $Z \to \eta_b + \gamma $ up to order ${\cal O}(\alpha_{s} v^2)$.}
\label{tb.etab}
\end{table}

\begin{table}[htb]
\begin{tabular}{c c c c c c}
\hline\hline
$\alpha_s(\mu_R)$  &  ${\rm LO}$  & ${\cal O}(\alpha_s)$ & ${\cal O}(v^2)$ & ${\cal O}(\alpha_{s} v^2)$ & {\rm Total} \\
\hline
$\alpha_s(2 m_b) =0.178$  & $146.24 $   & $-15.50 $  & $0.20$ & $0.19$ & $131.13  $ \\
$\alpha_s({m_{z}}/{2}) =0.132$  & $146.24 $   & $-11.50 $  & $0.20$ & $0.14$ & $135.08 $\\
$\alpha_s(m_{z}) =0.118$ & $146.24 $   & $-10.28 $  & $0.20$ & $0.13$ & $136.29 $\\
\hline\hline
\end{tabular}
\caption{Total and separate decay widths (in unit: eV) of $Z \to \Upsilon + \gamma $ up to order ${\cal O}(\alpha_{s} v^2)$.}
\label{tb.Upsilon}
\end{table}

The decay widths of the processes $Z \to H+\gamma$ are given in Tables \ref{tb.etac}, \ref{tb.Jpsi}, \ref{tb.etab} and \ref{tb.Upsilon}, where the contributions from different orders are shown explicitly\footnote{To obtain the contributions from different orders (in $v^2$ and $\alpha_s$) of the decay widths, we expand the quakonium mass $m_H$ appearing in Eqs.(\ref{eq.nrqcdfact}) and (\ref{eq.gamma}) by using the Gremm-Kapustin relation \cite{Gremm:1997dq}, i.e., $m_H^2 \approx 4m_Q^2(1+\langle v^2\rangle_{H})$.}. Results for three typical choices of renormalization scales ($2m_Q$, $m_{_Z}/2$ and $m_{_Z}$) are adopted in deriving those tables, while the factorization scale $\mu_{\Lambda}$ is taken as $m_Q$.

Tables \ref{tb.etac}, \ref{tb.Jpsi}, \ref{tb.etab} and \ref{tb.Upsilon} show that the total decay width of $Z \to J/\psi(\Upsilon)+\gamma$ is larger than that of $Z \to \eta_c(\eta_b)+\gamma$. The main reason is that the production of $J/\psi(\Upsilon)+\gamma$ is through the axial-vector coupling of $Z$ boson, while the production of $\eta_c(\eta_b)+\gamma$ is through the vector coupling of $Z$ boson, and the axial-vector coupling is stronger than the vector coupling in the $Zc\bar{c}\; (Zb\bar{b})$ interaction. The results also show that the QCD and relativistic corrections are important in these decay processes. More explicitly, when the scale of $\alpha_s$ is set as $m_{_Z}/2$, the ${\cal O}(\alpha_s)$, ${\cal O}(v^2)$ and ${\cal O}(\alpha_s v^2)$ corrections are $10.4\%$, $-18.9\%$ and $-9.4\%$ of the LO decay width for the $\eta_c$ case, which become $2.7\%$, $-3.7\%$ and $-4.3\%$ for the $J/\psi$ case, $-0.6\%$, $0.8\%$ and $0.2\%$ for the $\eta_b$ case, and $-7.9\%$, $0.1\%$ and $0.1\%$ for the $\Upsilon$ case, respectively. 

Tables \ref{tb.etac} and \ref{tb.Jpsi} show that for the $J/\psi$ case, the ${\cal O}(\alpha_s v^2)$ correction is larger than the ${\cal O}(\alpha_s)$ correction, and for the $\eta_c$ case, the ${\cal O}(\alpha_s v^2)$ correction is smaller but comparable with the ${\cal O}(\alpha_s)$ correction. Therefore, in those two cases, the ${\cal O}(\alpha_s v^2)$ corrections are important for the accurate theoretical predictions. Tables \ref{tb.etab} and \ref{tb.Upsilon} show that the ${\cal O}(\alpha_s v^2)$ corrections are relatively smaller for the $\eta_b$ and $\Upsilon$ cases. The main reason is that the value of $\langle v^2\rangle_{\Upsilon}$ obtained in Ref.\cite{Chung:2010vz} is small. If we take the values of $\langle v^2\rangle_{\eta_b}$ and $\langle v^2\rangle_{\Upsilon}$ as the average values of $v^2$ of a heavy quark in quarkonia calculated by using the relativistic Bethe-Salpeter equation, i.e., $0.0708$ for $\langle v^2\rangle_{\eta_b}$ and $0.0715$ for $\langle v^2\rangle_{\Upsilon}$~\cite{Wang:2020zbr}, the ${\cal O}(\alpha_s v^2)$ corrections which be increased to $-1.8\%$ and $-0.8\%$ of the LO decay widths for the $\eta_b$ and $\Upsilon$ cases, respectively.



\subsection{Uncertainty analysis}

In this subsection, we present an estimation on the theoretical uncertainties for those decay widths. The main uncertainty sources for those decay widths include the renormalization scale, the heavy quark masses, and the NRQCD LDMEs. For the uncertainties caused by different choices of renormalization scale, we estimate them by varying its magnitude between two energy scales $2 m_Q$ and $m_{_Z}$ with the central value $m_{_Z}/2$. Then the uncertainties caused by the renormalization scale can be read from Tables \ref{tb.etac}, \ref{tb.Jpsi}, \ref{tb.etab} and \ref{tb.Upsilon}. For the uncertainties caused by different choices of heavy quark masses, we estimate them by taking $m_c=1.4\pm 0.1 \,{\rm GeV}$ for the decays $Z \to \eta_c+ \gamma$ and $Z \to J/\psi+ \gamma$, and $m_b=4.6\pm 0.1 \,{\rm GeV}$ for the decays $Z \to \eta_b+ \gamma$ and $Z \to \Upsilon+ \gamma$. For the uncertainties caused by the LDMEs, we estimate them by varying $\langle {\cal O}_{1}\rangle_H$ and $\langle v^2\rangle_H$ with the ranges given in Eqs.(\ref{eq.ldmes1}, \ref{eq.ldmes2}, \ref{eq.ldmes3}). Then, the uncertainties caused by these uncertainty sources are
\begin{eqnarray}
&&\Gamma_{Z \to \eta_c + \gamma}=32.90^{+0.40 +2.73 +12.81 +4.91 } _{-0.04 -2.36 -11.98 -6.16} \,\, {\rm eV},\\
&&\Gamma_{Z \to J/\psi + \gamma}=261.02^{+0.46 +21.70 +39.75 +8.64} _{-4.27 -18.71 -32.63 -10.41} \,\, {\rm eV}, \\
&&\Gamma_{Z \to \eta_b + \gamma}=69.55^{+0.03 +1.73 +4.69 +0.23} _{-0.10 -1.65 -4.31  -0.23} \,\, {\rm eV}, \\
&&\Gamma_{Z \to \Upsilon + \gamma}=135.08^{+1.20 +3.29+9.11+0.11} _{-3.96-3.14-8.36-0.11} \,\, {\rm eV},
\end{eqnarray}
where the first uncertainty is caused by the renormalization scale, the second uncertainty is caused by the heavy quark mass, the third uncertainty is caused by $\langle {\cal O}_{1}\rangle_H$, and the last uncertainty is caused by $\langle v^2\rangle_H$. One may observe that the largest uncertainty is caused by the LDMEs $\langle {\cal O}_{1}\rangle_H$. By adding those uncertainties in quadrature, we obtain the total theoretical uncertainties for these decay widths, e.g.,
\begin{eqnarray}
&&\Gamma_{Z \to \eta_c + \gamma}=32.90 ^{+14.00} _{-13.67} \,\, {\rm eV}, \\
&&\Gamma_{Z \to J/\psi + \gamma}=261.02 ^{+46.10 } _{-39.26 } \,\, {\rm eV}, \\
&&\Gamma_{Z \to \eta_b + \gamma}=69.55^{+5.00} _{-4.62 } \,\, {\rm eV}, \\
&&\Gamma_{Z \to \Upsilon + \gamma}=135.08 ^{+9.76 } _{-9.77 } \,\, {\rm eV}.
\end{eqnarray}

Having the decay widths, the branching fractions for these decay processes can be calculated by using the total decay width of $Z$ boson, e.g. $\Gamma_{_Z}=2.4955\,{\rm GeV}$, which is taken from PDG~\cite{ParticleDataGroup:2022pth}. We obtain~\footnote{The uncertainty of $\Gamma_{_Z}$ given by the PDG is $0.0023{\rm GeV}$ \cite{ParticleDataGroup:2022pth}, which is only about $0.1\%$ of $\Gamma_{_Z}$. The uncertainties for the branching fractions caused by $\Gamma_{_Z}$ are only about $0.1\%$ of their central values. Thus the uncertainties caused by the errors of $\Gamma_{_Z}$ are negligible, we will neglect their contributions in the calculation of branching fractions.}
\begin{eqnarray}
&&{\rm Br}(Z \to \eta_c + \gamma)=1.32 ^{+0.56} _{-0.54}\times 10^{-8}, \\
&&{\rm Br}(Z \to J/\psi + \gamma)=1.04 ^{+0.18 } _{-0.16 }\times 10^{-7}, \\
&&{\rm Br}(Z \to \eta_b + \gamma)=2.79^{+0.20} _{-0.19 }\times 10^{-8}, \\
&&{\rm Br}(Z \to \Upsilon + \gamma)=5.41 ^{+0.39 } _{-0.39 }\times 10^{-8}.
\label{eq.Br}
\end{eqnarray}

The latest experimental upper limits on the two branching fractions issued by the ATLAS collaboration are ${\rm Br}(Z \to J/\psi + \gamma)<1.2 \times 10^{-6}$ and ${\rm Br}(Z \to \Upsilon + \gamma)<1.1\times 10^{-6}$ \cite{ATLAS:2022rej}. The branching fractions shown in Eq.(\ref{eq.Br}) are compatible with these upper limits.

\section{Summary}
\label{sec5}

We have calculated the decay widths for the rare $Z$-boson decays $Z \to H+ \gamma$ up to ${\cal O}(\alpha_{s} v^2)$ corrections, where $H =\eta_{c}$, $J/\psi$, $\eta_{b}$, and $\Upsilon$, respectively. Our calculations contain the NLO QCD correction ${\cal O}(\alpha_s v^0)$, the relativistic correction ${\cal O}(\alpha_s^0 v^2)$ and their joint correction ${\cal O}(\alpha_s v^2)$. Our numerical results show that these corrections are sizable and important, especially for the charmonium case. For examples, the ${\cal O}(\alpha_s v^2)$ correction is larger than ${\cal O}(\alpha_s v^0)$ correction for the $J/\psi+ \gamma$ case, and the ${\cal O}(\alpha_s v^2)$ correction is comparable with the ${\cal O}(\alpha_s v^0)$ correction for the $\eta_c+ \gamma$ case. For the case of $\eta_b+ \gamma$ and $\Upsilon+ \gamma$, the expansion over $\alpha_s$ and $v^2$ becomes more convergent than the charmonium case due to larger typical energy scale ${\cal O}(2m_b) >> {\cal O}(2m_c)$.

By taking the mentioned error sources into consideration, the obtained branching fractions are ${\rm Br}(Z \to \eta_c + \gamma)=1.32 ^{+0.56} _{-0.54}\times 10^{-8}$, ${\rm Br}(Z \to J/\psi + \gamma)=1.04 ^{+0.18 } _{-0.16 }\times 10^{-7}$, ${\rm Br}(Z \to \eta_b + \gamma)=2.79^{+0.20} _{-0.19 }\times 10^{-8}$ and ${\rm Br}(Z \to \Upsilon + \gamma)=5.41 ^{+0.39 } _{-0.39 }\times 10^{-8}$, respectively. The results for the charmonium case are compatible with the latest upper limits obtained by the ATLAS collaboration~\cite{ATLAS:2022rej}. The $e^+e^-$ colliders proposed in recent years, e.g., ILC, CEPC, FCC-ee, and Super $Z$ factory, provide new opportunities for studying these rare decays. For instance, about $7 \times 10^{11}$ $Z$ boson events can be generated at the CEPC due to $Z$ boson resonance effect~\cite{CEPCStudyGroup:2018ghi}. According to the branching fractions obtained in this work, we estimate that about $9\times 10^3$, $7\times 10^4$, $2\times 10^4$, $4\times 10^4$ events for $Z \to \eta_c+ \gamma$, $Z \to J/\psi+ \gamma$, $Z \to \eta_b+ \gamma$, $Z \to \Upsilon+ \gamma$ are expected to be generated at the CEPC. One may expect that these rare $Z$-boson decays can be studied at the $Z$-pole mode of these $e^+e^-$ colliders.

\hspace{2cm}

\noindent {\bf Acknowledgments:} This work was supported in part by the Natural Science Foundation of China under Grants No. 12005028, No. 12175025 and No. 12347101, by the Fundamental Research Funds for the Central Universities under Grant No. 2020CQJQY-Z003, by the Chongqing Natural Science Foundation under Grant No. CSTB2022NSCQ-MSX0415, and by the Chongqing Graduate Research and Innovation Foundation under Grant No. ydstd1912.

\hspace{2cm}

\end{document}